\documentclass{jpp}
\usepackage{graphicx}
\usepackage[dvipsname]{xcolor}
\usepackage{amsmath,amsfonts}
\usepackage{subfigure}
\usepackage{bm}
\usepackage[outdir=./]{epstopdf}
\usepackage[colorlinks=true, linkcolor=olive,citecolor=olive]{hyperref}

\shorttitle{Rotating Alfv{\'e}n waves in rotating plasmas}
\shortauthor{J.-M. Rax, R. Gueroult and N. J. Fisch}

\title{Rotating Alfv{\'e}n waves in rotating plasmas}

\author{J.-M. Rax\aff{1,2},  R. Gueroult\aff{3}
  \and N. J. Fisch\aff{4}}

\affiliation{\aff{1}Andlinger Center for Energy + the Environment, Princeton
University, Princeton, NJ 08540, USA
\aff{2}IJCLab, Universit\'{e} de Paris-Saclay, 91405 Orsay, France
\aff{3}LAPLACE, Universit\'{e} de Toulouse, CNRS, INPT, UPS, 31062 Toulouse, France
\aff{4}Department of Astrophysical Sciences, Princeton University, Princeton NJ 08540, USA}

\begin{document}

\maketitle

\begin{abstract}
Angular momentum coupling between a rotating magnetized plasma and torsional Alfv{\'e}n waves carrying orbital angular momentum (OAM) is examined. It is not only demonstrated that rotation is the source of Fresnel-Faraday rotation - or orbital Faraday rotation effects - for OAM carrying Alfv{\'e}n waves, but also that angular momentum from an OAM carrying Alfv{\'e}n wave can be transferred to a rotating plasma through the inverse process. For the direct process, the transverse structure angular rotation frequency is derived by considering the dispersion relation for modes with opposite OAM content. For the inverse process, the torque exerted on the plasma is derived as a function of wave and plasma parameters.
\end{abstract}

\section{Introduction}

Understanding the effect of rotation on plasma dynamics is essential to a wide range of applications. Besides original efforts motivated by microwave generation in magnetrons~\citep{Brillouin1945}, it has indeed been shown that rotation could enable new approches to thermonuclear confinement~\citep{Rax2017,Wilcox1959,Bekhtenev1980,Ochs2017b,Hassam1997,Fetterman2010,Fetterman2008}. Rotation has also be found to hold promise for developing plasma mass separation applications~\citep{Gueroult2017b,Gueroult2019}, either in pulsed plasma centrifuges~\citep{Bonnevier1966,Krishnan1981} or in steady-state cross-field rotating plasmas~\citep{Ohkawa2002,Shinohara2007,Gueroult2014,Fetterman2011,Gueroult2014}, advanced accelerators~\citep{Janes1965,Janes1965a,Janes1966,Thaury2013,Rax2010} and thrusters~\citep{Gueroult2013a}. But understanding the effect of rotation on plasma dynamics is also essential in a number of environments. Rotation is for instance key to the structure and stability of a number of astrophysical objects~\citep{Kulsrud1999,Miesch2009}. In light of this ubiquitousness, and because plasma waves are widely used both for control and diagnostics in plasmas, it seems desirable to understand what the effect of rotation on wave propagation in plasmas may be~\citep{Gueroult2023}. In fact the importance of this task was long recognised in geophysics and astrophysics, leading to extensive studies of low frequency MHD waves in rotating plasmas~\citep{Lehnert1954,Hide1969,Acheson1972,Acheson1973,Campos2010}, and notably of Alfv{\'e}n waves~\citep{Stix1992}. 

Meanwhile, following the discovery that electromagnetic waves carry both spin and orbital angular momentum~\citep{Allen1992,Allen2016,Andrews2012}, there have been numerous theoretical developments on spin-orbit interactions~\citep{Bliokh2015} in modern optics, which we note are now being applied to plasmas~\citep{Bliokh2022}. For spin angular momentum (SAM) carrying waves, that is circularly polarised waves, propagation through a rotating medium is known to lead to a phase-shift between eigenmodes with opposite SAM content~\citep{Player1976,Gueroult2019a,Gueroult2020}. This phase-shift is then the source of a rotation of polarization or polarization drag~\citep{Jones1976}, as originally postulated by Thomson~\citep{Thomson1885} and Fermi~\citep{Fermi1923}.  For orbital angular momentum (OAM) carrying waves, propagation through a rotating medium is the source of a phase-shift between eigenmodes with opposite OAM content~\citep{Goette2007}, leading to image rotation or Faraday-Fresnel Rotation (FFR)~\citep{Padgett2006}. 


This azimuthal Fresnel drag of OAM carrying waves, which can be viewed as an orbital Faraday rotation of the amplitude, was first derived~\citep{WisniewskiBarker2014} and observed~\citep{Franke-Arnold2011} in isotropic, nongyrotropic media. In contrast, propagation of OAM carrying wave in a rotating anisotropic (gyrotropic) medium poses greater difficulty since the polarization state and the wave vector direction - which are independent parameters for a given wave frequency in an isotropic medium - become coupled. Yet, it was recently shown that Faraday-Fresnel Rotation (FFR) is also found for the high frequency magnetized plasma modes that are Whistler-Helicon and Trivelpiece-Gould modes~\citep{Rax2021}. For such high frequency modes it was found that the main modifications induced by the plasma rotation are associated with Doppler shift and Coriolis effect in the dispersion
relation. Interestingly, we note that the result that rotation is the source of an azimuthal component for the group velocity of low frequency waves in magnetized plasmas when $\bm{\Omega}\cdot\mathbf{k}\neq 0$ was already pointed out in geophysics and astrophysics~\citep{Acheson1973}, but the connection to a Faraday-Fresnel rotation of the transverse structure of the wave did not seem to have been made. An added complexity for these low frequency modes is that one must, in addition to anisotropy and gyrotropy, consider the strong coupling to the inertial mode~\citep{Lighthill1980} that then comes into play. Revisiting this problem, we derive here in this study the expression for FFR for low frequency rotating Alfv{\'e}n waves in a rotating magnetized plasma.

This paper is organised as follows. After briefly recalling the configuration of interest and previous results in the next section, we construct in section~\ref{Sec:III} the spectrum of low frequency, small amplitude, fluid waves in a magnetized rotating plasma. The set of linearised Euler and Maxwell equations describes an oscillating Beltrami flow-force free field~\citep{Chandrasekhar1956} whose components are expressed with a cylindrical Chandrasekhar-Kendall (CK) potential~\citep{Chandrasekhar1957,Yoshida1991}. Then, in section~\ref{Sec:IV}, these orbital angular momentum carrying waves are shown to display a FFR under the influence of the plasma rotation. Section~\ref{Sec:V} focuses on the inverse problem when the orbital angular momentum of the wave is absorbed by the plasma. We derive in this case the torque exerted by this wave on the fluid driven as a function of the wave and plasma parameters. Finally section~\ref{Sec:VI} summarises the main findings of this study.

\section{Background}
\label{Sec:II}

In this study we consider a rotating magnetized plasma column with angular velocity $\bm{\Omega }=\Omega \mathbf{e}_{z}$ and static uniform magnetic field  $\mathbf{B}_{0}=B_{0}\mathbf{e_{z}}$. We write $(r,\theta,z)$ and $(x,y,z)$ cylindrical and Cartesian coordinates on cylindrical $\left(\mathbf{e}_{r},\mathbf{e}_{\theta},\mathbf{e}_{z}\right) $ and Cartesian $\left( \mathbf{e}_{r},\mathbf{e}_{\theta },\mathbf{e}_{z}\right)$ basis, respectively. The plasma dynamics is described assuming an inviscid and incompressible fluid model. We classically define the Alfv{\'e}n velocity $\mathbf{V}\doteq\mathbf{B}_{0}/\sqrt{\mu_{0}\rho }$ where $\mu_{0}$ is the permittivity of vacuum and $\rho$ the mass density of the fluid.

\begin{figure}
\begin{center}
\includegraphics[width=8cm]{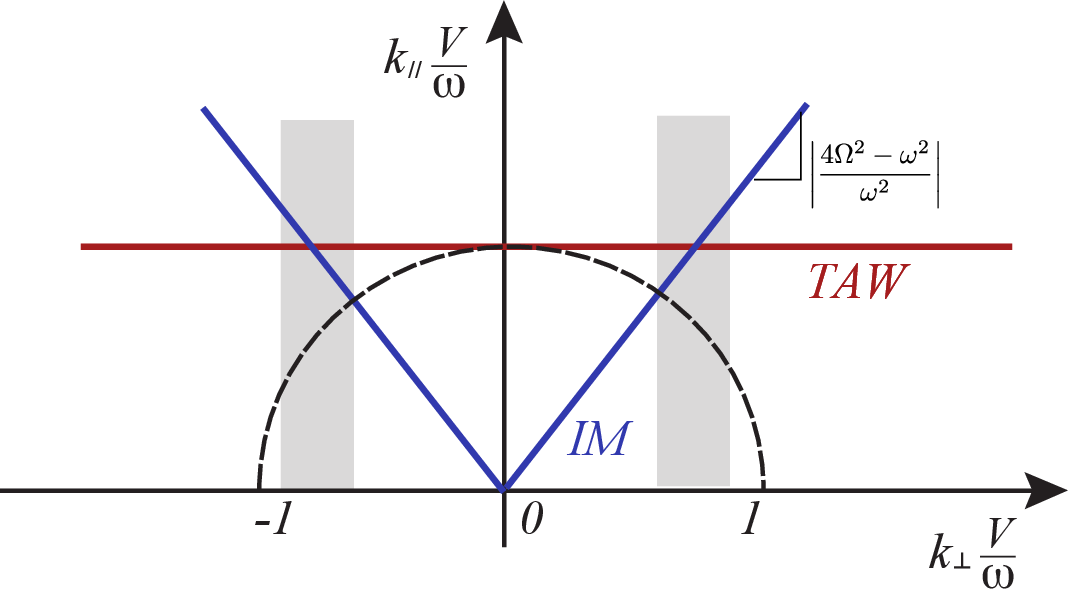}
\caption{Uncoupled dispersion of torsional Alfv{\'e}n waves (TAW) obtained for $B_{0}\neq 0$ and $\Omega =0$, and of inertial waves (IM) obtained for $\Omega \neq 0$ and $B_{0}=0$.}
\label{Fig:Fig1}
\end{center}
\end{figure}

In the simple case where $B_{0}=0$ and $\Omega \neq 0$ the rotating plasma behaves as an
ordinary rotating fluid and inertial waves can propagate. Taking a phase factor $\exp j\left( \omega t-k_{\Vert }z-k_{\perp }y\right) $, the dispersion relation for this inertial mode (IM) is~\citep{Lighthill1980}
\begin{equation}
\omega =\pm 2\Omega k_{\Vert }/\sqrt{k_{\Vert }^{2}+k_{\perp }^{2}}.
\end{equation}
Conversely, in the case where $\Omega =0$ but $B_{0}\neq 0$, Alfv{\'e}n waves can propagate in the magnetized plasma at rest. The dispersion of this torsional mode (TAW) is~\citep{Stix1992}
\begin{equation}
\omega =\pm B_{0}k_{\Vert }/\sqrt{\mu _{0}\rho }=\pm k_{\Vert }V.
\end{equation}
Note that compressional Alfv{\'e}n wave (CAW) are not considered here as we are considering
an incompressible plasma. The dispersion of uncoupled TAW and IM is plotted in Fig.~\ref{Fig:Fig1} in the $\left( k_{\Vert }V/\omega ,k_{\perp }V/\omega \right) $ plane for a given frequency $\omega $. In this figure the grey zones indicate regions of strong coupling between TAW and IM. Note that we have normalised for convenience the wave-vector to $\omega /V$, and that even for the unmagnetized IM branch.

In the more general case where both $B_{0}\neq 0$ and $\Omega \neq 0$, then a strong coupling between IM and TAW modes rearranges the spectrum and gives rise to two new branches~\citep{Lehnert1954,Acheson1973}. Since as already pointed out by \cite{Acheson1973} the group velocity of these modes for waves such that $\bm{\Omega}\cdot\mathbf{k}\neq0$ has an azimuthal  component, then we expect Fresnel-Faraday Rotation as recently identified for Trivelpiece-Gould and Whistler-Helicon high frequency electronic modes~\citep{Rax2021}.


\section{Rotating Alfv{\'e}n waves in a rotating plasma}
\label{Sec:III}

In this section we examine the properties of low frequency waves carrying orbital angular momentum in a rotating magnetized plasma. 

\subsection{Classical modes}

Two methods can be used to identify and describe the coupling between the angular momentum of a rotating plasma column and the angular momentum of a wave propagating in this rotating magnetized plasma. One is to consider the transformation laws of the various parameters from the lab frame to a rotating frame. The other is to perform the study in the lab frame starting from first principles. Here we will use the first method, similarly to original contributions on MHD waves in rotating conductive fluids~\citep{Lehnert1954,Hide1969}, and solve the perfect MHD dynamics to calculate the rotating plasma linear response for the low frequency branches where the coupling between the fields and the particles is large. By working in the co-rotating frame (R) rather than in the laboratory frame (L), both the Coriolis force $2\bm{\Omega}\times \mathbf{v}$ and the centrifugal forces $\bm{-\nabla }\psi$ with $\psi =-\Omega ^{2}r^{2}/2$ must be taken into account.

We model the evolution of the wave velocity field $\mathbf{v}\left(\mathbf{r},t\right)$ using Euler's equation under the assumption of zero viscosity
\begin{equation}
\frac{\partial \mathbf{v}}{\partial t}+\left( \mathbf{v}\cdot \bm{\nabla}\right) \mathbf{v}+2\bm{\Omega }\times \mathbf{v} =-\bm{\nabla}\left( \frac{P}{\rho }+\psi \right) +\frac{1}{\mu _{0}\rho }\left( \bm{\nabla}\times \mathbf{B}\right) \times \left( \mathbf{B+B}_{0}\right),  \label{euler}
\end{equation}
and the evolution of the wave magnetic field $\mathbf{B}\left(\mathbf{r},t\right)$ using Maxwell-Faraday's equation under the assumption of perfect conductivity
\begin{equation}
\frac{\partial \mathbf{B}}{\partial t} =\bm{\nabla}\times \left[\mathbf{v}\times \left( \mathbf{B+B}_{0}\right) \right],\label{MaxF}
\end{equation}
where $\rho $ is the mass density of the fluid and $P$ is the pressure. These dynamical relations are completed by the flux conservation law
\begin{equation}
\bm{\nabla }\cdot \mathbf{B}=0
\end{equation} 
for the magnetic field and the incompressibility relation
\begin{equation}
\bm{\nabla}\cdot \mathbf{v}=0
\end{equation}
for the velocity field. As already mentioned this last relation will restrict the plasma behaviour to the
Alfv{\'e}nic dynamics associated with torsional waves.

We then consider a small amplitude magnetohydrodynamic perturbation, propagating along and around the $z$ axis, described by a magnetic perturbation 
\begin{equation}
\mathbf{B}\left( r,\theta ,z,t\right) = \mathfrak{B}\left( r,\theta ,z\right)\exp (j\omega t)
\end{equation}
with respect to the uniform static magnetic field $\mathbf{B}_{0}$ = $B_{0}\mathbf{e}_{z}$. The wave frequency $\omega $ is assumed smaller than the ion cyclotron frequency and larger than the collision frequency to validate the use of the perfect MHD model  Eqs.(\ref{euler}, \ref{MaxF}). The oscillating magnetic wave $\mathbf{B}$ is associated with an oscillating hydrodynamic velocity perturbation $\mathbf{v}$ 
\begin{equation}
\mathbf{v}\left( r,\theta ,z,t\right) =\mathbf{u}\left( r,\theta ,z\right)\exp (j\omega t)\text{,}
\end{equation}
with respect to rotating frame velocity equilibrium $\mathbf{v}_{0}=\mathbf{0}$. The pressure $P$ balances the centrifugal force at equilibrium $\bm{\nabla}\left( P+\rho \psi \right) =\mathbf{0}$ and the pressure perturbation is $p\left( r,\theta ,z\right)\exp j\omega t$. To first order in these perturbation the linearisation of Eqs.~\eqref{euler} and \eqref{MaxF} gives
\begin{gather}
j\omega \mathbf{u}+2\bm{\Omega }\times \mathbf{u} =-\bm{\nabla}\left( p/\rho \right) +\frac{1}{\mu _{0}\rho }\left( \bm{\nabla}\times 
\mathfrak{B}\right) \times \mathbf{B}_{0},  \label{lin1} \\
j\omega  \mathfrak{B} =\left( \mathbf{B}_{0}\cdot \bm{\nabla}\right)\mathbf{u}.  \label{lin2}
\end{gather}
Flux conservation and incompressibility provide the two additional conditions 
\begin{gather}
\bm{\nabla}\cdot \mathbf{u}=0\text{, }\\
\bm{\nabla}\cdot \mathfrak{B}=0\text{.}  \label{lin3}
\end{gather}
Taking the curl of both Eqs.~(\ref{lin1}, \ref{lin2}) and eliminating $\mathfrak{B}$ gives a linear relation for the velocity perturbation
\begin{equation}
\omega ^{2}\bm{\nabla}\times \mathbf{u}+2j\omega \left( \bm{\Omega}
\cdot \bm{\nabla}\right) \mathbf{u}+\left( \mathbf{V}\cdot \bm{\nabla}\right) ^{2}\left( \bm{\nabla}\times \mathbf{u}\right) =\mathbf{0}\text{.}  \label{linu}
\end{equation}

\begin{figure}
\begin{center}
\includegraphics[width=8cm]{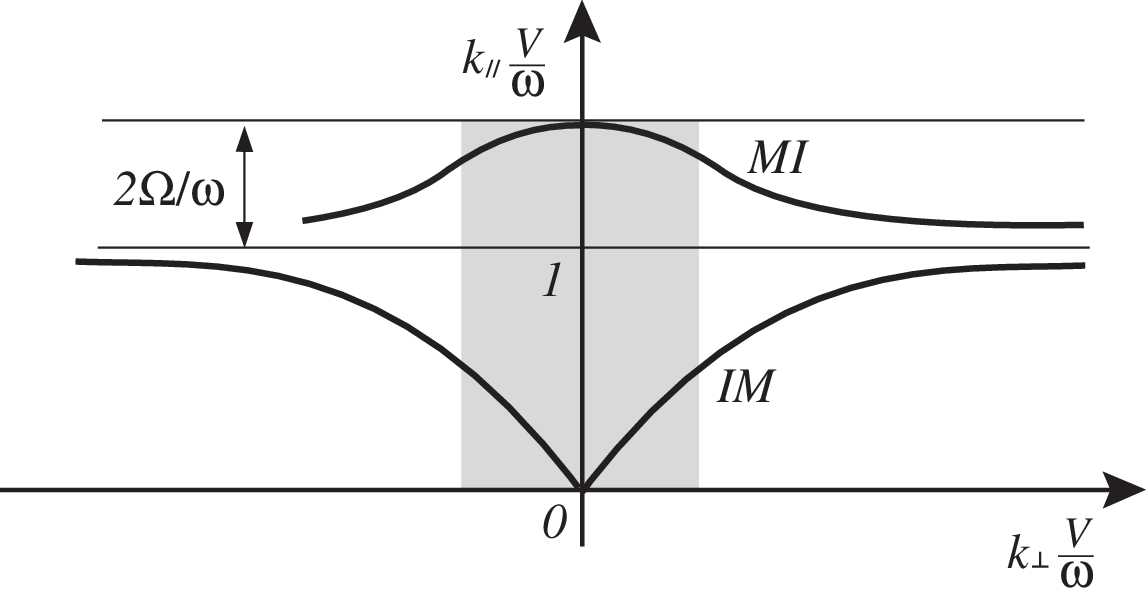}
\caption{Coupled dispersion of magnetoinertial waves (MI) and inertial waves (IM).}
\label{Fig:Fig2}
\end{center}
\end{figure}

Now if ones Fourier analyses this velocity perturbation as a superposition of plane waves
\begin{equation}
\mathbf{u}\left( \mathbf{r}\right) \exp j\omega t=\exp [j\left( \omega t-%
\mathbf{k}\cdot \mathbf{r}\right)] \text{,}  \label{plane}
\end{equation}
that is to say put the emphasis on the linear momentum dynamics rather than on the angular momentum one, one recovers the two branches of Alfvenic/Inertial perturbations in a rotating plasma~\citep{Lehnert1954,Acheson1973}. Specifically, plugging Eq.~\eqref
{plane} into Eq.~\eqref{linu} and then taking the cross product $j\mathbf{k}\times $ of this algebraic relation one obtain the dispersion relation
\begin{equation}
\omega ^{2}-\left( \mathbf{k}\cdot \mathbf{V}\right) ^{2}=\pm 2\omega \left( 
\bm{\Omega} \cdot \mathbf{k}\right) /\left| \mathbf{k}\right| \text{.}
\label{disp1}
\end{equation}
These two branches, which are illustrated in Fig.~\ref{Fig:Fig2}, have been widely investigated within the context of geophysical and astrophysical magnetohydrodynamics models. For short wavelengths the $\Omega =0$ torsional Alfv{\'e}n wave (TAW) splits into inertial (IM) and a magneto-inertial (MI) waves. For long wavelengths, that is in the grey zone in Fig.~\ref{Fig:Fig2}, inertial terms dominate the dispersion and the IM mode is found to reduce to its zero rotation behaviour already shown in Fig.~\ref{Fig:Fig1}. Note finally that the torsional Alfv{\'e}n wave is recovered for large $k_{\perp }$ where a local dispersion becomes valid as opposed to small $k_{\perp }$ where the large wavelength allows the wave to probe the large scale behaviour of the rotation.

\subsection{Beltrami flow}



Instead of this usual procedure using a full Fourier decomposition as given by Eq.~(\ref{plane}), we start here by considering a travelling perturbations along $z$ of the form 
\begin{equation}
\mathbf{u}\left( r,\theta,z\right) =\mathbf{w}(r,\theta) \exp (-jk_{\Vert }z).
\label{Eq:OAM_pert}
\end{equation}
Note that this is analog to what was already done by~\cite{Shukla2012} to study OAM carrying dispersive shear Alfv{\'e}n waves though in this earlier study the paraxial approximation and a two fluid model were used, and the plasma was considered at rest (\emph{i.~e.} non-rotating). Plugging Eq.~\eqref{Eq:OAM_pert} in the dispersion relation for a rotating plasma Eq.~\eqref{linu} gives
\begin{equation}
\bm{\nabla}\times \mathbf{u}=\mathcal{K}\mathbf{u} \label{eig7}
\end{equation}
where we have defined
\begin{equation}
\mathcal{K}\left( k_{\Vert },\omega \right)\doteq2\frac{\Omega }{\omega }k_{\Vert }\left( \frac{k_{\Vert }^{2}V^{2}}{\omega ^{2}}-1\right) ^{-1}\text{.}  \label{eig2}
\end{equation}
From Eq.~\eqref{lin2} the oscillating magnetic field then writes 
\begin{equation}
\omega \mathfrak{B}=-\sqrt{\mu _{0}\rho }k_{\Vert }V\mathbf{u}.  \label{eig 8}
\end{equation}

The two modes identified in Fig.~\ref{Fig:Fig2} can be recovered from Eq.~\eqref{eig2}. More specifically, for $k_{\Vert }V>\omega $ Eq.~\eqref{eig7} describes an Alfv{\'e}n wave modified by inertial effect. Conversely for $k_{\Vert }V<\omega $ Eq.~\eqref{eig7} describes an inertial wave modified by MHD coupling. In the following we will focus on the Alfv{\'e}n wave dynamics and thus assume $\mathcal{K}>0$. 

Equation~\eqref{eig7} is characteristic of a \textit{Beltrami }flow~\citep{Chandrasekhar1956}. As such $\mathbf{u}$ can be written in terms of the so called \textit{Chandrasekhar-Kendall }(CK) potential $\Phi$~\citep{Chandrasekhar1957} as
\begin{align}
\mathbf{u} & =\frac{1}{\mathcal{K}}\bm{\nabla}\times\left(\bm{\nabla}\times\Phi\mathbf{e}_{z}\right)+\bm{\nabla}\times\Phi\mathbf{e}_{z}\nonumber\\
 & = -\left[\frac{1}{\mathcal{K}}\bm{\nabla}\times\mathbf{e}_{z}\times\bm{\nabla}+\mathbf{e}_{z}\times\bm{\nabla}\right]\Phi  \label{helm2}
\end{align}
where the CK potential is solution of the scalar Helmholtz equation 
\begin{equation}
\Delta \Phi +\mathcal{K}^{2}\Phi =0\text{.}  \label{helm9}
\end{equation}
One verifies that the three components of Eq.~\eqref{helm2} are independent.

Before examining the structure of OAM carrying modes through the CK potential, two additional results can be obtained from Eq.~\eqref{eig2}. First, for the Fourier decomposition used above, plugging Eq.~\eqref{disp1} in Eq.~\eqref{eig2} gives
\begin{equation}
\frac{\mathcal{K}^{2}}{k_{\Vert }^{2}} =1+\frac{k_{\perp }^{2}}{k_{\Vert }^{2}}>1.  \label{dispomeg1}
\end{equation}
Second, we can derive the dimensionless group-velocity dispersion coefficient 
\begin{equation}
\frac{\omega }{\mathcal{K}}\frac{\partial \mathcal{K}}{\partial \omega } =-\frac{k_{\Vert }}{\mathcal{K}}\frac{\partial \mathcal{K}}{\partial k_{\Vert }}=\frac{k_{\Vert }^{2}V^{2}+\omega ^{2}}{k_{\Vert }^{2}V^{2}-\omega ^{2}}  \label{dispomeg3}
\end{equation}
which we will use later to explicit the axial wave vector difference for two eigenmodes with opposite OAM content.

\subsection{Structure of OAM carrying modes}

Because we are interested in waves carrying orbital angular momentum around $z$ and linear momentum along $z$, we search for solutions of the form 
\begin{equation}
\Phi \left( r,\theta ,z\right) =\phi \left( r\right) \exp [-j\left( m\theta+k_{\Vert }z\right)]  \label{helm3}
\end{equation}
where $m\in\Bbb{Z}$ is the azimuthal mode number associated with the orbital angular momentum of the wave. From Eq.~\eqref{helm9} the radial amplitude of this rotating CK potential $\phi(r)$ must be solution of the Bessel equation
\begin{equation}
\frac{1}{r}\frac{d}{dr}\left( r\frac{d\phi }{dr}\right) -\frac{m^{2}}{r^{2}}\phi +\left( \mathcal{K}^{2}-k_{\Vert }^{2}\right) \phi =0\text{.}  \label{helm1}
\end{equation}
Since as shown in Eq.~\eqref{dispomeg1} $\mathcal{K}^{2}>k_{\Vert }^{2}$,  $\phi(r)$ is in general the combination of Bessel functions of the first and the second kind and order $m\in \Bbb{Z}$, $J_{m}$ and $Y_{m}$. Yet, the finite value of $\phi $ at $r=0$ demands to restrict the physical solution to Bessel functions of the first kind $J_{m}$ so that we find
\begin{equation}
\phi \left( r\right) =J_{m}\left( \alpha r\right) \label{Bess1}
\end{equation}
with the cylindrical dispersion relation 
\begin{equation}
\alpha ^{2}+k_{\Vert }^{2}=\mathcal{K}^{2}\left( k_{\Vert },\omega \right) \text{.}
\label{Eq:disp_alpha}
\end{equation}
Note that, like the ordinary plane wave Eq. (\ref{plane}) used in the standard analysis, the cylindrical Bessel waves Eq.~\eqref{Bess1} can not be normalised.

Putting these pieces together one finally gets 
\begin{align}
\mathbf{v}& =\left[\frac{1}{\mathcal{K}}\bm{\nabla}\times\mathbf{e}_{z}\times\bm{\nabla}+\mathbf{e}_{z}\times\bm{\nabla}\right]
 J_{m}\left( \sqrt{\mathcal{K}^{2}-k_{\Vert }^{2}}r\right) \exp[
j\left( \omega t-m\theta -k_{\Vert }z\right)]\nonumber\\
& =-\frac{\omega \mathbf{B}}{\sqrt{\mu _{0}\rho }k_{\Vert }V} \text{.}  \label{vel2}
\end{align}
The components in the plasma frame of a rotating Alfv{\'e}n wave with azimuthal mode number $m$ thus have an amplitude proportional to combination of Bessel functions of the first kind and of orders $m$ and $m\pm 1$. All these Bessel functions have the same radial dependence, namely $\sqrt{\mathcal{K} ^{2}\left( k_{\Vert },\omega \right)-k_{\Vert}^{2}}r$, where $\mathcal{K}\left( k_{\Vert },\omega \right)$ is given by Eq.~\eqref
{eig2}.

\section{Direct rotational Fresnel drag-orbital Faraday rotation}
\label{Sec:IV}

Let us now rewrite these perturbations as seen from the laboratory frame. We use the index $R$ for the rotating plasma rest frame and $L$ for the laboratory frame. The radial Eulerian coordinates $r$ and $z$ are unchanged through this change of frame or reference, but azimuthal coordinates $\theta$ changes, with
\begin{gather}
\left. r \right| _{L}=\left. r\right| _{R}\\
\left. z \right| _{L}=\left. z\right| _{R}\\
\left. \theta \right| _{L}=\left.\theta \right| _{R}+\Omega t.
\end{gather}
Since the axial wave-vector is unchanged $\left.k_{\Vert }\right |_{R}=\left. k_{\Vert }\right |_{L}$, the phase of the wave in the plasma rest-frame 
\begin{equation}
\omega t-k_{\Vert }z\pm m\left. \theta \right |
_{R}
\end{equation}
 becomes
 \begin{equation}
 \left( \omega \mp m\Omega \right) t-k_{\Vert }z\pm m\left. \theta \right |_{L}
 \end{equation}
 in the laboratory frame.

\begin{figure}
\begin{center}
\includegraphics[width=10cm]{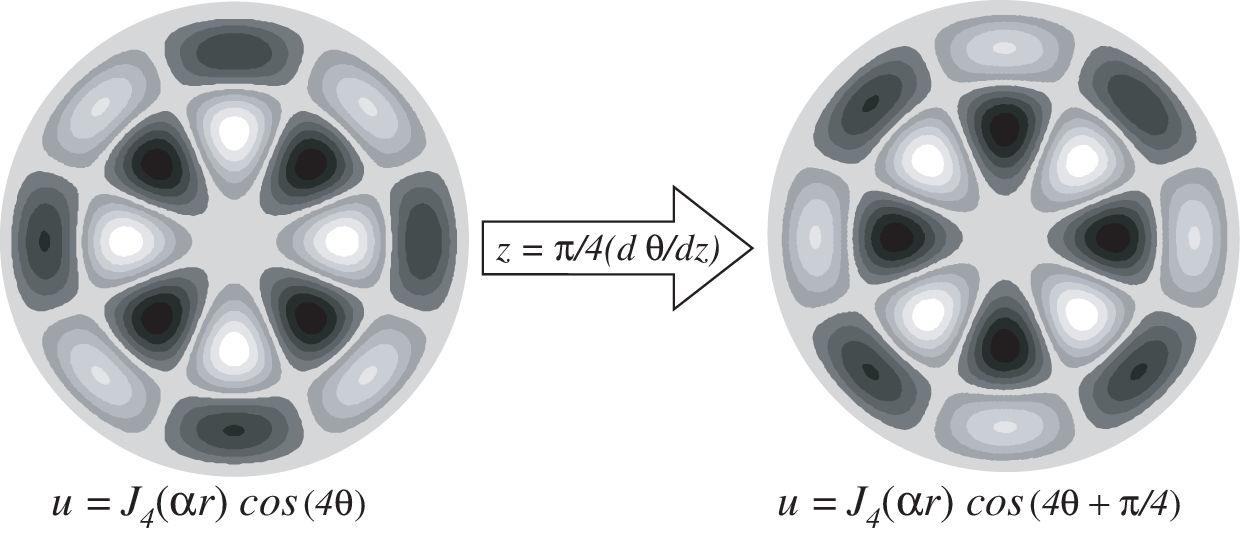}
\caption{Fresnel drag-Faraday rotation of the Chandrasekhar-Kendall potential describing an
Alfv{\'e}n-Beltrami wave with $m=\pm 4$ after a propagation along a path $z=\pi/4\left( d\theta /dz\right)$.}
\label{Fig:Fig3}
\end{center}
\end{figure}

Equipped with these transformations we can now describe the conditions to observe Fresnel-Faraday Rotation. For this we consider two CK potentials  describing two Alfv{\'e}n modes with opposite OAM content in the rotating frame $R$
\begin{gather}
\left. \Phi _{+}\right| _{R} =J_{m}\left( \alpha r\right) \exp \left(j\left[\left( \omega -m\Omega \right) t-\left( k_{\Vert }-\delta k_{\Vert }\right)z-m\left. \theta \right| _{R}\right]\right),  \label{ffr1} \\
\left. \Phi _{-}\right| _{R} =J_{-m}\left( \alpha r\right) \exp \left(j\left[\left( \omega +m\Omega \right) t-\left( k_{\Vert }+\delta k_{\Vert }\right)z+m\left. \theta \right| _{R}\right]\right).  \nonumber
\end{gather}
These transform in the CK potentials in the laboratory frame $L$
\begin{gather}
\left. \Phi _{+}\right| _{L} =J_{m}\left( \alpha r\right) \exp \left(j\left[\omega t-\left( k_{\Vert }-\delta k_{\Vert }\right) z-m\left. \theta \right|_{L}\right]\right),  \label{ffl} \\
\left. \Phi _{-}\right| _{L} =J_{-m}\left( \alpha r\right) \exp \left(j\left[\omega t-\left( k_{\Vert }+\delta k_{\Vert }\right) z+m\left. \theta \right|_{L}\right]\right),  \nonumber
\end{gather}
as a result of the rotational Doppler shift $\left. \theta \right |_{L}=\left. \theta \right | _{R}+\Omega t$. These Alfv{\'e}n CK potentials $\left. \Phi _{\pm }\right| _{L}$ can be driven by a multicoil antenna similar to that used to study Whistler-Helicon modes~ \citep{Stenzel2014,Stenzel2015,Stenzel2015a,Stenzel2015b,Urrutia2015,Urrutia2016,Stenzel2018,Stenzel2019}. The radial field pattern is then a superposition of $+m$ and $-m$ Bessel amplitudes $J_{\pm
m}\left( \alpha r\right) $ where $\alpha $ is associated with the radial modulation of the antenna currents~\citep{Rax2021}. The antenna then sets both the radial wave-vector $\alpha$ and the frequency $\omega$, whereas the axial wave-vectors $k_{\Vert }\pm \delta k_{\Vert }$ are solutions of the rotating frame dispersion relation. From Eq.~\eqref{Eq:disp_alpha}
\begin{gather}
\alpha  =\sqrt{\mathcal{K}^{2}\left( k_{\Vert }-\delta k_{\Vert },\omega -m\Omega
\right) -\left( k_{\Vert }-\delta k_{\Vert }\right) ^{2}}\text{,}
\label{eq1} \\
\alpha  =\sqrt{\mathcal{K}^{2}\left( k_{\Vert }+\delta k_{\Vert },\omega +m\Omega
\right) -\left( k_{\Vert }+\delta k_{\Vert }\right) ^{2}}\text{.}
\label{eq2}
\end{gather}
Since we assume $\omega \gg \Omega $ and $k_{\Vert }\gg \delta k_{\Vert }$ we can Taylor
expand Eq.~\eqref{eq1} and Eq.~\eqref{eq2} to get $\delta
k_{\Vert }$, leading to
\begin{equation}
\frac{k_{\Vert }}{\mathcal{K}}\delta k_{\Vert }=\frac{\delta k_{\Vert }}{2}\frac{\partial
\mathcal{K}\left( \omega ,k_{\Vert }\right) }{\partial k_{\Vert }}+\frac{m\Omega}{2} \frac{%
\partial \mathcal{K}\left( \omega ,k_{\Vert }\right) }{\partial \omega }\text{.}
\end{equation}
Eq.~\eqref{dispomeg3} can then be used to finally write the axial wave-vector difference $\delta k_{\Vert }$ for two modes with the same frequency $\omega $, the same radial amplitude $\left| J_{m}\left( \alpha r\right) \right|$ and equal but opposite azimuthal number $\left| m\right|$ as  
\begin{equation}
\frac{\delta k_{\Vert }}{k_{\Vert }}=\frac{1}{2}m\frac{\Omega }{\omega }\frac{1+\frac{k_{\Vert }^{2}V^{2}}{\omega ^{2}}}{1-\frac{k_{\Vert }^{2}}{\mathcal{K}^{2}} +\left( 1+\frac{k_{\Vert }^{2}}{\mathcal{K}^{2}}\right) \frac{k_{\Vert
}^{2}V^{2}}{\omega ^{2}}}  \label{ffr2}
\end{equation}
where $\mathcal{K}\left( k_{\Vert },\omega ,\Omega \right) $ is given by Eq.~\eqref{eig2}. This implies that there will be a difference in the axial phase velocity $\omega /\left( k_{\Vert}\pm \delta k_{\Vert }\right) $ of these two modes, and because these two modes rotates in opposite direction due to their opposite azimuthal mode number, the transverse structure of the sum of these modes will rotate. This is Fresnel drag-Faraday orbital rotation effects. Specifically, if one launches a wave which is a superposition of $+m$ and $-m$ modes such that at the antenna location $z=0$
\begin{equation}
\left. \Phi \right| _{z=0}=J_{m}\left( \alpha r\right) \left( \exp [j\left(
\omega t-m\left. \theta \right| _{L}\right)] +\left( -1\right) ^{m}\exp
[j\left( \omega t+m\left. \theta \right| _{L}\right)] \right) \text{,} 
\nonumber
\end{equation}
the wave transverse amplitude rotates as it propagates along $z>0$ with an angular velocity along the propagation axis 
\begin{equation}
\left. \frac{d\theta }{dz}\right| _{L}=\frac{\delta k}{m}=\frac{1}{2}\frac{\Omega }{\omega }k_{\Vert }\mathcal{K}^{2}%
\frac{k_{\Vert }^{2}V^{2}+\omega ^{2}}{k_{\Vert }^{2}V^{2}\left(
\mathcal{K}^{2}+k_{\Vert }^{2}\right) +\omega ^{2}\left( \mathcal{K}^{2}-k_{\Vert }^{2}\right) }
\text{.}  \label{rot}
\end{equation}
This CK potential rotation is illustrated in Fig.~\ref{Fig:Fig3} for the case $m=4$. Eqs.~(\ref{ffr2}, \ref{rot}) quantifies the direct Faraday-Fresnel effect of Alfv{\'e}n waves in rotating plasmas, completing the similar results previously obtained for Trivelpiece-Gould and Whistler-Helicon modes~\citep{Rax2021}. The $1/m$ factor in Eq.~\eqref{rot} comes from the fact that the image constructed from the superposition of $\pm m$ modes has a $2m$-fold symmetry.

To conclude this section it was shown that besides the Fresnel-Faraday Rotation associated to a phase velocity difference for $m$ and $-m$ modes, there can also be a spliting of the
envelope of a $\left( m,-m\right) $ wave packet if the
group velocity for co-rotating ($m$) and counter-rotating ($-m$) modes were different~\citep{Rax2021}. We note that this second effect is also present here for Alfv{\'e}n waves in rotating plasmas. Indeed, given a radial wave-vector $\alpha $ the dispersion relation is $\mathcal{D}=\mathcal{K}^{2}-k_{\Vert }^{2}-\alpha ^{2}=0$, so that from Eq. (\ref{dispomeg3}) the axial group velocity is given by 
\begin{equation}
-\frac{\partial \mathcal{D}}{\partial k_{\Vert }}/\frac{\partial \mathcal{D}}{\partial \omega }=\frac{k_{\Vert }}{\mathcal{K}\partial \mathcal{K}/\partial \omega }-\frac{\omega }{k_{\Vert }}
\text{.}  \label{group}
\end{equation}
and one verifies from Eq.~\eqref{eig2} that the group velocity for a mode $(k_{\Vert }+ \delta k_{\Vert },m)$ and that for a mode $(k_{\Vert }- \delta k_{\Vert },-m)$ are different. Rather than deriving here an explicit formula for the Fresnel-Faraday splitting, we consider in the next section the inverse Fresnel-Faraday effect associated with wave absorption.

\section{Inverse rotational Fresnel drag and angular moment absorption}
\label{Sec:V}

In a perfectly conducting inviscid plasma there is no power absorption. The power exchange between the oscillating electromagnetic field and the plasma is purely reactive. To obtain an irreversible (active) angular momentum absorption, on needs a dissipative mechanism. Two different wave orbital angular momentum absorption mechanisms can be considered. One is resonant collisionless absorption, the other is collisional absorption. The former was recently studied in \cite{Rax2023} through quasilinear theory and will not be considered here. Instead, we consider in this section a weakly dissipative
plasma where the ideal MHD hypothesis of perfect conductivity is relaxed and the inviscid assumption of zero viscosity no longer apply. In both case, collisionnal or collisionless, each time an energy $\delta \mathcal{U}$ is absorbed by the plasma, an axial angular momentum $\delta L=m\delta \mathcal{U}/\omega $ is also absorbed by the plasma~\citep{Rax2017,Rax2023}. The rate of decay of the wave angular momentum is hence equal to the wave induced density of torque on the plasma $\Gamma =dL/dt$. In steady-state, this angular momentum transfer $d\Gamma/dt$ is balanced by viscous damping of the velocity shear and Ohmic dissipation of the radial charge polarisation sustaining the rotation. This dissipation is larger in the collisionnal case considered here than in the collisionless regime considered in \cite{Rax2023}.

Specifically, we introduce two dissipative collisional coupling to our dissipation-less system Eqs.~(\ref{lin1}, \ref{lin2}), namely finite viscosity $\rho \mu $ and finite resistivity $\mu _{0}\eta$. We
follow the notation of \cite{Taylor1989} (devoted to Alfv{\'e}n wave helicity absorption) and introduce the magnetic diffusion coefficient $\eta $ and the kinematic viscosity $\mu $. Ohm's law then writes $\mathbf{E}+\mathbf{v\times B}=\mu _{0}\eta \mathbf{j}$ and the system Eqs.~(\ref{lin1}, \ref{lin2}) becomes
\begin{gather}
j\omega \mathbf{u}+2\bm{\Omega }\times \mathbf{u} =-\bm{\nabla}\left( p/\rho \right) +\frac{1}{\mu _{0}\rho }\left( \bm{\nabla}\times\mathfrak{B}\right) \times \mathbf{B}_{0}+\mu \Delta \mathbf{u},
\label{col1} \\
j\omega \mathfrak{B}=\left( \mathbf{B}_{0}\cdot \bm{\nabla}\right)\mathbf{u}+\eta \Delta\mathfrak{B}. \label{col2}
\end{gather}
Since we assume weak dissipation, the resistive term $\eta \Delta \mathbf{B}$ in Maxwell-Faraday's equation and the viscous term $\mu \Delta \mathbf{u}$ in Navier-Stokes equation can be evaluated with the dispersive properties of the non dissipative dispersion relation. Within the bounds of this perturbative expansion scheme ($\mathcal{K}^{2}\eta \ll \omega $ and $\mathcal{K}^{2}\mu \ll \omega $), and for the perturbation $\mathbf{u}\left( r,\theta,z\right) =\mathbf{w}(r,\theta) \exp (-jk_{\Vert }z)$ already given in Eq.~\eqref{Eq:OAM_pert}, we get from Eqs.~(\ref{eig7}, \ref{eig 8}) the non-dissipative Laplacians
\begin{gather}
\Delta \mathbf{u} = -\mathcal{K}^{2}\mathbf{u}, \\
\Delta \mathfrak{B}= -\mathcal{K}^{2}\mathfrak{B}.
\end{gather}
Plugging these results into Eqs.~(\ref{col1}, \ref{col2}) yields the system
\begin{gather}
j\omega \mathbf{u}+2\bm{\Omega }\times \mathbf{u} =-\bm{\nabla}\left( p/\rho \right) +\frac{1}{\mu _{0}\rho }\left( \bm{\nabla}\times \mathfrak{B}\right) \times \mathbf{B}_{0}-\mathcal{K}^{2}\mu \mathbf{u}, \\
j\omega \mathfrak{B}=\left( \mathbf{B}_{0}\cdot \bm{\nabla}\right) \mathbf{u}-\mathcal{K}^{2}\eta \mathfrak{B}
\end{gather}
where now viscous and resistive dissipation introduce a local relaxation.

We then take the rotational of the first equation and eliminate $\mathfrak{B}$ using the second equation to get 
\begin{equation}
\left[ \left( j\omega +\mathcal{K}^{2}\mu \right) \left( j\omega +\mathcal{K}^{2}\eta \right) \right] \bm{\nabla}\times \mathbf{u}+2j\left( j\omega +\mathcal{K}^{2}\eta\right) k_{\Vert }\Omega \mathbf{u}+k_{\Vert }^{2}V^{2}\bm{\nabla}\times \mathbf{u=0}\text{.}
\end{equation}
After some algebra we find that the linearised dissipative regime of velocity and field low frequency oscillations is now described by 
\begin{gather}
\bm{\nabla}\times \mathbf{u} =\left[ \mathcal{K}_{R}\left( k_{\Vert },\omega \right)-j\mathcal{K}_{I}\left( k_{\Vert },\omega \right)\right] \mathbf{u}
\label{eig12} \\
\left( \omega -j\mathcal{K}^{2}\eta \right) \mathbf{B} =-\sqrt{\mu _{0}\rho }k_{\Vert }V\mathbf{u}
\end{gather}
rather than by the collisionles Eqs. (\ref{eig7}, \ref{eig 8}), where we have defined the two real wave-vectors $\mathcal{K}_{R}\approx \mathcal{K}\gg \mathcal{K}_{I}$ through 
\begin{equation}
\mathcal{K}_{R}\left( k_{\Vert },\omega \right) -j\mathcal{K}_{I}\left( k_{\Vert },\omega \right) =2\Omega \frac{\left( \omega -j\mathcal{K}^{2}\eta \right) k_{\Vert }}{k_{\Vert }^{2}V^{2}-\left(\omega -j\mathcal{K}^{2}\mu \right) \left( \omega -j\mathcal{K}^{2}\eta \right) }\text{.}
\label{KIKI}
\end{equation}

We then consider an initial value problem with a weakly decaying wave of the form 
\begin{equation}
\mathbf{v}= \mathbf{u}\exp \left[j\left( \omega +j\nu \right) t\right]
\end{equation}
with $\omega \gg \nu $, and with the structure 
\begin{equation}
\mathbf{v}=\left[\frac{1}{\mathcal{K}_{R}-j\mathcal{K}_{I}}\bm{\nabla}\times\mathbf{e}_{z}\times\bm{\nabla}+\mathbf{e}_{z}\times\bm{\nabla}\right]J_{m}\left( \alpha r\right) \exp \left(j\left[ \left( \omega +j\nu \right)
t-m\theta -k_{\Vert }z\right] \right)
\end{equation}
where $\alpha $ is a real number, $\omega $ and $k_{\Vert }$ are given, and the damping rate $\nu \left( \omega ,k_{\Vert },\mathcal{K}\right) $ is to be determined from the weak dissipation expansion of the dispersion relation 
\begin{equation}
\alpha ^{2}+k_{\Vert }^{2}=\left[ \mathcal{K}_{R}\left(  k_{\Vert },\omega +j\nu \right)-j\mathcal{K}_{I}\left(  k_{\Vert },\omega +j\nu \right) \right] ^{2}
\end{equation}
obtained by plugging this solution in Eq.~\eqref{eig12}. Taylor expanding this last relation for $\nu\ll\omega $, the lowest order real part gives the collisionless dispersion
\begin{equation}
\alpha ^{2}\left(  k_{\Vert },\omega \right)  =\mathcal{K}_{R}^{2}\left(  k_{\Vert },\omega \right)-k_{\Vert}^{2}\approx \mathcal{K}^{2}\left(  k_{\Vert },\omega \right)-k_{\Vert }^{2}  \label{tor2}
\end{equation}
while the lowest order imaginary part gives a relation for the decay rate $\nu $
\begin{equation}
\nu\left(  k_{\Vert },\omega \right) \frac{\partial \mathcal{K}_{R}\left( \omega\right) }{\partial \omega } =\mathcal{K}_{I}\left(  k_{\Vert },\omega \right) \approx \frac{\mathcal{K}^{3}}{\omega }\left[\eta+ (\mu +\eta)\left({\frac{k_{\Vert
}^{2}V^{2}}{\omega ^{2}}-1}\right)^{-1}\right].   \label{tor3}
\end{equation}
Here we took $\partial \mathcal{K}_{R}/\partial \omega \approx \partial \mathcal{K}/\partial \omega $ and used Eq. (\ref{dispomeg3}).

Finally, Eq.~\eqref{tor3} can be used to write an equation for the evolution of the wave energy density $\mathcal{U}$ 
\begin{equation}
\frac{d\mathcal{U}}{dt}=-2\nu \mathcal{U}=-2\mathcal{K}_{I}\left( \frac{\partial \mathcal{K}_{R}}{\partial \omega }\right) ^{-1}\mathcal{U}.
\label{Eq:ener_evol}
\end{equation}
For a rotating Alfv{\'e}n wave, this energy density $\mathcal{U}$ has three distinct components 
\begin{equation}
\mathcal{U}=\frac{\left\langle B^{2}\right\rangle }{2\mu _{0}}+\frac{\varepsilon _{0}}{2}\left\langle \left( \mathbf{v}\times \mathbf{B}_{0}\right)^{2}\right\rangle +\frac{\rho }{2}\left\langle v^{2}\right\rangle 
\label{Eq:ener}
\end{equation}
where $\left\langle {}\right\rangle $ indicate an average over the fast $\omega $ oscillations. The first term on the right hand side is the magnetic energy, the second term is the electric energy and the third term is the kinetic energy. This energy density can be rewritten using the Alfv{\'e}n velocity $V$ and the velocity of light $c$ as 
\begin{equation}
\mathcal{U}=\frac{\rho }{2}\left[ \left\langle \mathbf{v}^{2}\right\rangle \left( 1+%
\frac{V^{2}}{c^{2}}\left( 1+\frac{k_{\Vert }^{2}c^{2}}{\omega ^{2}}\right)
\right) -\left\langle \left( \mathbf{v}\cdot \frac{\mathbf{V}}{c}\right)
^{2}\right\rangle \right].   \label{vel}
\end{equation}
Combining Eq.~\eqref{Eq:ener_evol}, Eq.~\eqref{Eq:ener} and the relation between energy and angular momentum absorption, one finally gets
\begin{equation}
\Gamma =2\rho \frac{m}{\omega }\mathcal{K}_{I}\left( \frac{\partial \mathcal{K}_{R}}{\partial
\omega }\right) ^{-1}\left[ \left\langle \mathbf{v}^{2}\right\rangle \left(
1+\frac{V^{2}}{c^{2}}\left( 1+\frac{k_{\Vert }^{2}c^{2}}{\omega ^{2}}\right)
\right) -\frac{\left\langle \left( \mathbf{v}\cdot \mathbf{V}\right)
^{2}\right\rangle }{c^{2}}\right]   \label{tor}
\end{equation}
where $\mathcal{K}_{R}$ and $\mathcal{K}_{I}$ are given by Eq.~\eqref{KIKI} and $\mathbf{v}$ is given by Eq.~\eqref{vel2}. 


\section{Conclusion}
\label{Sec:VI}

Building on previous contibutions studying Alfv{\'e}n waves in rotating plasmas in geophysical and astrophysical settings, we have examined here the dynamics of orbital angular momentum  (OAM) carrying torsional Alfv{\'e}n waves in a rotating plasma. 
It is found that two new couplings between the orbital angular momentum of the Alfv{\'e}n waves and the angular momentum of the rotating plasma exist.

One is Fresnel-Faraday rotation (FFR), that is a rotation of the transverse structure of the wave due to the medium's rotation, which had already been predicted for the high frequency electronic modes that are Trivelpiece-Gould and Whistler-Helicon modes~\citep{Rax2021}. Extending these earlier contributions, direct Fresnel-Faraday rotation (FFR) for torsional Alfv{\'e}n waves in a rotating plasma is described by Eqs.~(\ref{ffr2}) and (\ref{rot}). It is the orbital angular momentum analog of the polarization drag effect for spin angular momentum waves~\citep{Jones1976,Player1976}. An important distinction found here though is that while rotation did not introduce new high frequency modes so that FFR for Trivelpiece-Gould and Whistler-Helicon modes was simply the consequence of the interplay between Coriolis force and rotational Doppler shift~\citep{Rax2021}, the strong coupling to the inertial mode that exists for Alfv{\'e}n waves in rotating plasmas complexifies this picture.

The second coupling is the inverse effect through which the OAM carrying wave exerts a torque on the plasma. Inverse FFR is described by Eqs.~(\ref{KIKI}) and (\ref{tor}). This inverse effect is akin to the spin angular momentum inverse Faraday effect but for the orbital angular momentum of the wave. It is found that for a plasma with non-zero collisional absorption the damping of an OAM carrying wave is the source of a torque on the plasma.





Looking ahead, these results suggest that direct FFR could in principle be used to diagnose plasma rotation with Alfv{\'e}n waves. Conversely, it may be possible to utilise inverse FFR to sustain plasma rotation through Alfv{\'e}n waves angular momentum absorption. The detailed analysis of these promising prospects is left for future studies.



\section*{Acknowledgments}
The authors would like to thank Dr. E. J. Kolmes, I. E. Ochs, M. E. Mlodik
and T. Rubin for constructive discussions.

\section*{Funding}
This work was supported by the U.S. Department of Energy (N. J. F.,  grant number ARPA-E Grant No. DE-AR001554); and the French National Research Agency (R. G., grand number ANR-21-CE30-0002). JMR acknowledges Princeton University and the Andlinger Center for Energy + the Environment for the ACEE fellowship which made this work possible.

\section*{Declaration of interests}
The authors report no conflict of interest.

\bibliographystyle{jpp}

\bibliography{JMR_Alfven}

\end{document}